\documentclass[12pt]{article}

\usepackage{amssymb}
\usepackage{subfigure}
\usepackage{color}
\usepackage{epsfig,amssymb,amsfonts,amsmath,graphicx,dsfont,cite,xfrac}
\usepackage{authblk}
\definecolor{mygray}{gray}{0.5}

\usepackage{cite}
\usepackage[colorlinks=true,linkcolor=blue,citecolor=red]{hyperref}
\newcommand{\be}{\begin{equation}}
\newcommand{\ee}{\end{equation}}
\newcommand{\bea}{\begin{eqnarray}}
\newcommand{\eea}{\end{eqnarray}}

\title{On the construction of non-Hermitian Hamiltonians with all-real spectra through supersymmetric algorithms}

\author[${1,2}$]{Kevin Zelaya}
\author[${3}$]{Sara Cruz y Cruz}
\author[${1}$]{Oscar Rosas-Ortiz}

\affil[${1}$]{\footnotesize Physics Department, Cinvestav, AP 14-740, 07000
M\'exico City, Mexico}

\affil[${2}$]{\footnotesize Centre de Recherches Math\'ematiques, Universit\'e de Montr\'eal, Montr\'eal, Qu\'ebec H3C 3J7, Canada}

\affil[${3}$]{\footnotesize Instituto Polit\'ecnico Nacional, UPIITA, Av I.P.N 2580, C.P. 07340, M\'exico City, Mexico
}

\date{}
\begin{document}

\maketitle

\begin{abstract}
The energy spectra of two different quantum systems are paired through supersymmetric algorithms. One of the systems is Hermitian and the other is characterized by a complex-valued potential, both of them with only real eigenvalues in their spectrum. The superpotential that links these systems is complex-valued, parameterized by the solutions of the Ermakov equation, and may be expressed either in nonlinear form or as the logarithmic derivative of a properly chosen complex-valued function. The non-Hermitian systems can be constructed to be either parity-time-symmetric or non-parity-time-symmetric.
\end{abstract}

%%%%%%%%%%%%%%%%%%%%%%%%%%%%%%%%%%%%%%%%%%%%%

%---------------------------------------> Section
\section{Introduction}

The supersymmetric formulation of quantum mechanics is a subject of intense activity in contemporary physics. It is addressed to analyze the spectral properties of exactly solvable potentials as well as to construct new integrable quantum models \cite{Bag00,Coo01,Mie04}. Sustained by the factorization method \cite{Mie84,And84a}, the supersymmetric approach is basically algebraic \cite{And84b} and permits the pairing between the spectrum of a given (well-known) Hamiltonian $H_0$ to the spectrum of a second (generally unknown) Hamiltonian $H_1$. In terms of differential operators, it has been found that the factorization of either $H_0$ or $H_1$ is not unique \cite{Mie84} and that the pairing of $H_0$ with $H_1$ is ruled by a Darboux transformation \cite{And84c}, which was introduced in 1882 \cite{Dar82} (see historical details in e.g. \cite{Mie04,Ros99}). The keystone is a solution $u$ (not necessarily normalizable) of the eigenvalue equation $H_0 u= \epsilon u$ that is used to generate the Darboux transformation $V_1(x) = V_0(x) + 2 \frac{d}{dx} \beta(x)$ \cite{Mie84,And84a}, where $\beta(x) = -\frac{d}{dx} \ln u(x)$ is called superpotential and $\epsilon$ the factorization energy. Remarkably, not only Hermitian but also non-Hermitian Hamiltonians $H_1$ can be produced as supersymmetric partners of a given exactly solvable (either Hermitian or non-Hermitian) Hamiltonian $H_0$. Indeed, depending on the properties of $V_0(x)$ and $\beta(x)$, the new potential $V_1(x)$ may be either real or complex-valued. In any case, the spectrum of the new Hamiltonian $H_1$ includes either all-real eigenvalues or a combination of real and complex eigenvalues, see e.g. \cite{Bay96,Can98,And99,Bag01,Ros03,Ros07,Fer08a,Fer08b,Mir13,Ros15,Jai17,Ros18,Gar19,Bag19}.

Quite recently, a complex-valued superpotential defined by the nonlinear expression
\begin{equation}
\beta(x)= -\frac{d}{dx} \ln \alpha'(x) +i\frac{\lambda}{\alpha^{2}(x)}, \quad \lambda \in \mathbb R, 
\label{eq:LD1}
\end{equation}
has been provided to produce new classes of non-Hermitian Hamiltonians $H_1$ with all-real spectra \cite{Ros15}. The function $\alpha(x)$ is a solution of the Ermakov equation \cite{Erm80}:
\begin{equation}
-\frac{d^2}{dx^2} \alpha(x) + V_{0} (x) \alpha(x) = \epsilon \alpha(x) + \frac{\lambda^{2}}{\alpha^{3}(x)},
\label{eq:FM6}
\end{equation}
which is reduced to the eigenvalue equation $H_0 \alpha = \epsilon \alpha$ for $\lambda =0$. The eigenfunctions of the resulting non-Hermitian Hamiltonians $H_1$ satisfy some properties of interlacing of zeros that permit the study of the related systems as if they were Hermitian \cite{Jai17}. Indeed, a bi-orthogonal basis can be introduced to facilitate the construction of coherent states for such a class of systems \cite{Ros18}. Moreover, the factorization energy $\epsilon$ can be positioned at any arbitrary position in the spectrum of $H_1$ \cite{Gar19}. Notedly, the eigenvalues of the non-Hermitian Hamiltonians $H_1$ are all-real regardless of whether $H_1$ is parity-time-symmetric \cite{Ben98} or not.

In this communication we briefly revisit the method developed in \cite{Ros15,Jai17,Ros18,Gar19} and show that the nonlinear superpotential (\ref{eq:LD1}) can be also expressed in the `canonical form' $\beta(x) = -\frac{d}{dx} \ln u(x)$, where $u$ is an eigenfunction of $H_0$ with very concrete profile. The results presented here generalize the approach introduced in \cite{Can98}, where it is guessed that a complex linear-combination of eigenfunctions of $H_0$ may be useful to construct complex-valued potentials $V_1(x)$. We provide a pair of examples where the new potentials are either parity-time-symmetric or non-parity-time-symmetric.

%---------------------------------------> Section
\section{Factorization method and non-Hermitian Hamiltonians}

Consider an initial Hamiltonian
\begin{equation}
H_{0}=-\frac{d^{2}}{dx^{2}}+V_{0}(x),
\label{eq:FM1} 
\end{equation}
with $V_{0}(x)$ a real-valued potential defined in $\mbox{Dom} V_0 \subseteq \mathbb R$. We assume that the energy eigenvalues $E^{(0)} \in\mathbb{R}$ and eigenfunctions $\phi(x)$ of the related eigenvalue equation $H_{0}\phi(x)=E^{(0)} \phi(x)$ are already known. In particular, the bounded solutions $\phi_n(x)$ belong to the discrete eigenvalues $E_n^{(0)}$, $n=0,1,\ldots$ Let us introduce a pair of non-mutually adjoint operators, $A$ and $B$, such that
\begin{equation}
H_{0}=AB+\epsilon, \quad A=-\frac{d}{dx}+\beta(x), \quad B=\frac{d}{dx}+\beta(x),
\label{eq:FM4}
\end{equation}
where $\beta(x)$ is in general a complex-valued function and $\epsilon$ is a real constant. After comparing (\ref{eq:FM4}) with (\ref{eq:FM1}) one arrives at the Riccati equation
\begin{equation}
-\beta'+\beta^{2}=V_{0}(x)-\epsilon, \quad \beta'=\frac{d\beta}{dx}.
\label{eq:FM5}
\end{equation} 
Provided a solution of (\ref{eq:FM5}), reversing the order of the factors in (\ref{eq:FM4}) gives 
\begin{equation}
H_{1}=BA+\epsilon=-\frac{d^{2}}{dx^{2}}+V_{1}(x), \quad V_{1}(x)
 = V_{0}(x)+2\beta'(x).
\label{eq:FM9}
\end{equation}
Notice that the new operator $H_1$ is not self-adjoint since $V_1$ is complex-valued in general. Indeed, $H_1^{\dagger} = A^{\dagger} B^{\dagger} + \epsilon = -\frac{d^2}{dx^2} + V_1^* \neq H_1$. Nevertheless, the pair $H_0$ and $H_1$ satisfies the intertwining relationships
\begin{equation}
BH_{0}=H_{1}B, \quad H_{0}A=AH_{1},
\label{eq:FM10}
\end{equation}
so that the eigenvalue equation $H_{1}\psi_{n}=E_{n}^{(1)}\psi_{n}$, $n=0,1,\ldots$, is automatically solved by the set
\begin{equation}
\psi_{n+1}=\frac{1}{\sqrt{E_{n}^{(0)}-\epsilon}}B\phi_{n} \, , \quad A\psi_{0}=0 \, , \quad E_{n+1}^{(1)}=E_{n}^{(0)} \, , \quad E_{0}^{(1)}=\epsilon \, .
\label{eq:FM12}
\end{equation}
The functions $\psi_n(x)$ are complex-valued and such that the zeros of their real and imaginary parts satisfy some theorems of interlacing \cite{Jai17}.

%---------------------------------------> Subsection
\subsection{Complex-valued potentials with all-real spectra}

In the conventional supersymmetric approaches the solution of the Riccati equation (\ref{eq:FM5}) is usually taken to be real-valued. However, complex-valued solutions are feasible even for real-valued potentials $V_0$ and real factorization energies $\epsilon$. Indeed, the real and imaginary parts of Eq.~(\ref{eq:FM5}) lead to a coupled system which is solved by the complex-valued superpotential (\ref{eq:LD1}). Assuming, with no loss of generality, that $\alpha(x)$ is real-valued, it may be shown that the solution of  the Ermakov (\ref{eq:FM6}) can be written as \cite{Ros15}:
\begin{equation}
\alpha(x)=\left[ a u_{1}^{2}(x)+b u_{1}(x)u_{2}(x)+c u_{2}^{2}(x) \right]^{1/2}, 
\label{eq:FM7}
\end{equation}
where $u_{1,2}$ are solutions of the system
\begin{equation}
-u_{1,2}''+V_{0}u_{1,2}=\epsilon u_{1,2}, \quad W(u_{1},u_{2})=u_{1}u_{2}'-u_{1}'u_{2}=W_{0},
\label{eq:FM8}
\end{equation}
with $W_0 = \mbox{const}$. The function $\alpha$ is free of zeros in $\mbox{Dom} V_0$ if the set $\{a,b,c \}$ is integrated by positive numbers that are constrained as follows 
\begin{equation}
b^{2}-4ac=-4\lambda^{2}/W_{0}^{2}.
\label{eq:FM8-1}
\end{equation} 
Using the superpotential (\ref{eq:LD1}), with $\alpha$ given in (\ref{eq:FM7}), the new potential (\ref{eq:FM9}) is now given by the nonlinear expression
\begin{equation}
V_1(x) = V_0(x) -2 \left( \ln \alpha(x) \right)'' + i  \left( \frac{2 \lambda}{\alpha^2(x)} \right)', \quad \lambda \in \mathbb R.
\label{pot1}
\end{equation}
Notice that the results of the conventional supersymmetric approaches \cite{Bag00,Coo01,Mie04} are automatically recovered for $\lambda=0$. On the other hand, it may be shown that the imaginary part of $V_1(x)$ satisfies the {\em condition of zero total area} \cite{Jai17}:
\begin{equation}
\int_{Dom V_0} \mbox{Im} V_1(x) dx = \left. \frac{2\lambda}{\alpha^2(x)} \right \vert_{Dom V_0} =0,
\label{zero}
\end{equation}
so that the total probability is conserved. The latter means that the potentials (\ref{pot1}) can be addressed to represent open quantum systems with balanced gain (acceptor) and loss (donor) profile \cite{Ele17}. 

%---------------------------------------> Subsubsection
\subsubsection{Parity-time-symmetric potentials} Potentials featuring the parity-time symmetry \cite{Ben98} represent a particular case of the applicability of the condition of zero total area (\ref{zero}). Such potentials are invariant under parity (P) and time‐reversal (T) transformations in quantum mechanics, so that a necessary condition for PT-symmetry is $V(x) = V^*(-x)$, where ${}^*$ stands for complex conjugation. For initial potentials $V_0(x)$ such that $V_0(x) = V_0(-x)$, one can show that making $b = 0$ in (\ref{eq:FM7}) is sufficient to get $V_1^*(x) = V^*_1(-x)$. In other words, the parity-time symmetry is a consequence of the condition of zero total area in our approach.

%---------------------------------------> Subsubsection
\subsubsection{Non-parity-time-symmetric potentials} For $V_0(x) \neq V_0(-x)$ the property $V_1^*(x) = V^*_1(-x)$ does not hold anymore, so the complex-valued potentials (\ref{pot1}) have all-real spectra although they are non-parity-symmetric. Diverse examples have been already discussed in e.g. \cite{Ros15,Jai17,Ros18,Gar19}. Quite recently the pseudo-Hermiticity and supersymmetric approaches  have been combined to get new classes of non-parity-time-symmetric potentials with all-real spectra \cite{Bag19}. Interestingly, such potentials can be manipulated to induce phase transitions where conjugate pairs of complex eigenvalues emerge in the spectrum. Similar results have been reported in \cite{Jai16}, where the condition of zero total area (\ref{zero}) plays a relevant role. The discussion on the subject is out of the scope of the present work and will be reported elsewhere.

%---------------------------------------> Subsection

\subsection{Recovering the canonical form of the superpotential}

We wonder if the nonlinear expression (\ref{eq:LD1}) can be reduced to the canonical form $\beta = -\frac{d}{dx} \ln u(x)$. Keeping this in mind, we first rewrite \eqref{eq:LD1} as 
\begin{equation}
\beta=-\frac{\frac{1}{2}(\alpha^{2})'-i\lambda}{\alpha^{2}}.
\label{eq:LD2}
\end{equation}
Using (\ref{eq:FM7}) and (\ref{eq:FM8-1}) we factorize the $\alpha$-function in the form
\begin{equation}
\alpha^{2}=\frac{1}{a} \left[ a u_{1}+ \left( \frac{b}{2}+i\frac{\lambda}{W_{0}} \right)u_{2} \right] \left[ a u_{1}+ \left( \frac{b}{2}-i\frac{\lambda}{W_{0}} \right)u_{2} \right].
\label{eq:LD3}
\end{equation}
In turn, expanding the numerator of Eq.~(\ref{eq:LD2}) yields
\begin{equation}
\frac{1}{2}\left(\alpha^{2}\right)'-i\lambda=a u_{1}u_{1}'+c u_{2}u_{2}'+b u_{1}'u_{2}+\left(\frac{bW_{0}}{2}-i\lambda \right),
\label{eq:LD4}
\end{equation}
where we have used the Wronskian defined in (\ref{eq:FM8}). The latter result is now factorized:
\begin{equation}
(C_{0}u_{1}'+C_{1}u_{2}')(D_{0}u_{1}+D_{1}u_{2}).
\label{eq:LD5}
\end{equation}
The coefficients $C_{0},C_{1},D_{0},D_{1}$ are defined by comparing the expanded version of \eqref{eq:LD5} with \eqref{eq:LD4}. One gets
\begin{equation}
\frac{1}{2}\left(\alpha^{2}\right)'-i\lambda=\frac{1}{a} \left[ a u_{1}'+ \left( \frac{b}{2}-i\frac{\lambda}{W_{0}} \right)u_{2}' \right] \left[ a u_{1}+ \left( \frac{b}{2}+i\frac{\lambda}{W_{0}} \right)u_{2} \right] \, .
\label{eq:LD6}
\end{equation}
Finally, the substitution of  \eqref{eq:LD3} and~\eqref{eq:LD6} into \eqref{eq:LD2} produces
\begin{equation}
\beta=  -\frac{\alpha'(x)}{\alpha(x)}+i\frac{\lambda}{\alpha^{2}(x)}
= -\frac{d}{dx}\ln \left[ a u_{1}+ \left( \frac{b}{2}-i\frac{\lambda}{W_{0}} \right)u_{2} \right].
\label{eq:LD7}
\end{equation}
Thus, the function we are looking for is given by the linear superposition
\begin{equation}
u = a u_{1}+ \left( \frac{b}{2}-i\frac{\lambda}{W_{0}} \right)u_{2},
\label{u}
\end{equation}
where the constants $a$, $b$ and $\lambda$ are linked by the condition (\ref{eq:FM8-1}). If $\lambda =0$ the constraint (\ref{eq:FM8-1}) becomes $b= \pm 2 \sqrt{ac}$, so that the coefficients of the superposition (\ref{u}) are real numbers, $u= \sqrt{a} \left( \sqrt{a} u_1 + \sqrt{c} u_2 \right)$, as expected. 

The expression (\ref{eq:LD7}) shows that the superpotential $\beta(x)$ can be written in either the nonlinear form (\ref{eq:LD1}), or as the logarithmic derivative of the function $u$ defined in (\ref{u}). The latter is a linear superposition of the solutions of (\ref{eq:FM8}) with complex coefficients that are uniquely defined by the condition (\ref{eq:FM8-1}). Notice that the derivation of the $u$-function (\ref{u}) generalizes the approach introduced in \cite{Can98}, where it is guessed that a linear combination of $u_{1,2}$ would give rise to complex-valued potentials $V_1$ whenever the appropriate complex coefficients have been included. As an example, in \cite{Can98} the authors provide the coefficients that produce a family of oscillator-like complex-valued potentials. They also apply their method to study the potential $V_1(x) = -\tfrac12 (ix)^N$, $N\geq 2$, introduced in \cite{Ben98}, and describe some other potentials that can be studied within their approach. However, no general rule to fix the appropriate complex coefficients is given in \cite{Can98}. In contrast, the linear superposition (\ref{u}) is general in the sense that the rule (\ref{eq:FM8-1}) applies for any differentiable and exactly solvable real-valued initial potential $V_0(x)$. Diverse examples have been already provided in \cite{Ros15,Jai17,Ros18,Gar19}.

%---------------------------------------> Section
\section{Examples and discussion of results}

\begin{figure}[htb]

\centering
\subfigure[$\mbox{Re} V_1(x)$]{\includegraphics[width=0.3\textwidth]{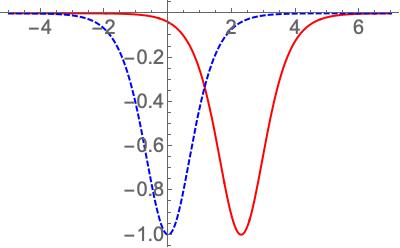} } 
\hskip1cm
\subfigure[$\mbox{Im} V_1(x)$]{\includegraphics[width=0.3\textwidth]{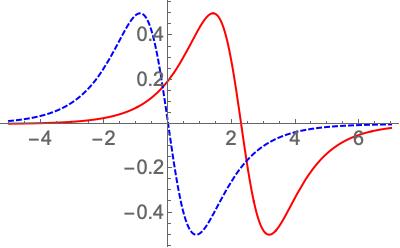} } 

\caption{\footnotesize 
The real and imaginary parts of the complex-valued potentials with all-real spectra (\ref{pot1}) derived from the expressions of the free particle provided in (\ref{pt}) with $b\neq0$, $a=1.5$ (red curve), and $b=0$, $a=1$ (dotted-blue curve). In all cases $\lambda=\kappa=1$.}
\label{Fig1}
\end{figure}

As immediate examples let us discuss the regular complex-valued potential $V_1(x)$ generated by the following initial potentials:

\vskip1ex
$\bullet$ {\bf Free particle.} Given $V_0(x)=0$, the basis set is $u_1= e^{ikx}$ and $u_2=e^{-ikx}$, with $W_0 = -2ik$. To get a real-valued $\alpha$-function we take $k= i \frac{\kappa}{2}$, with $\kappa >0$. Without loosing generality we now make $a=c$. Then,
\begin{equation}
\alpha (x) = \left[ 2 a \cosh \kappa x + b \right]^{1/2}, \qquad u(x)= a e^{-\kappa x/2} + \left( \frac{b}{2} - i\frac{\lambda}{\kappa} \right) e^{\kappa x/2}.
\label{pt}
\end{equation}
The potentials $V_1(x)$ are depicted in Fig.~\ref{Fig1}, they are of the P\"oschl-Teller type, generalize the well known family of regular (real-valued) supersymmetric partners of the free particle \cite{Mie00}, and  satisfy the condition of zero total area (\ref{zero}). These potentials include only one bound state of energy $E^{(0)}=-\tfrac14 \kappa^2$. The effect of $b\neq 0$ is to slide the potential to the right (red curve in Figure~\ref{Fig1}), so that $V_1(x)$ is parity-time-invariant after the appropriate shift. The latter is just because the initial potential $V_0(x)=0$ satisfies the condition $V_0(x) = V_0(-x)$ and exhibits, at the same time, translational symmetry $V_0(x) = V_0(x+x_0)$. One may say that, in the present case, the translational symmetry is invariant under the Darboux transformations (\ref{pot1}).

%%%%%%%%%%%%%
%%%%%%%%%%%%%

\begin{figure}[htb]

\centering
\subfigure[$\mbox{Re} V_1(x)$]{\includegraphics[width=0.3\textwidth]{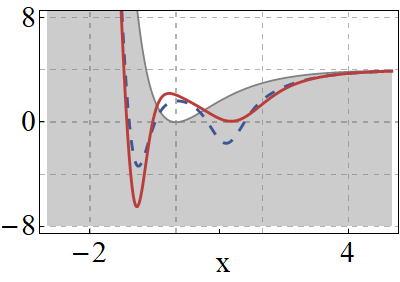} } 
\hskip1cm
\subfigure[$\mbox{Im} V_1(x)$]{\includegraphics[width=0.3\textwidth]{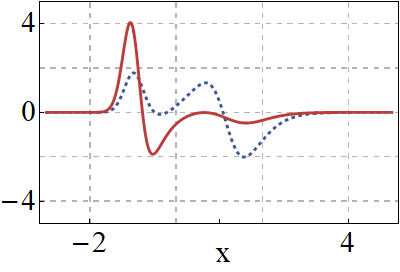} } 

\caption{\footnotesize 
The real and imaginary parts of the complex-valued potential with all-real spectra (\ref{pot1}) derived for the expressions of the Morse potential provided in (\ref{eq:MO2}) with $a=c=1$ (red curve), and $a=1$, $c=1/3$, $b=0$ (dotted-blue curve). In all cases $\lambda=2$ and $\epsilon=1$. The gray area delimitates the initial Morse potential.
}
\label{Fig3}
\end{figure}

%\vskip1ex
$\bullet$ {\bf Morse potential.} It is clear that the condition $V_0(x) = V_0(-x)$ cannot be applied on the Morse potential
\begin{equation}
V_0(x)=\Gamma_{0}(1-e^{-\gamma x})^{2}, \quad x\in\mathbb{R}.
\label{eq:MO1}
\end{equation}
Then, the potentials (\ref{pot1}) associated to (\ref{eq:MO1}) are non-parity-time-symmetric for any values of the set $\{a,b,c\}$. The condition $\Gamma_{0}>\gamma^{2}/2$ ensures that at least one bound state exists. It may be shown \cite{Gar19} that two linear independent solutions of~\eqref{eq:FM8} for $\epsilon\in\mathbb{R}$ are given in terms of confluent hypergeometric functions as follows
\begin{equation}
\begin{aligned}
& u_{1}(x)=e^{-y/2}y^{\sigma}{}_{1}F_{1}\left( \sigma+\frac{1}{2}-d;1+2\sigma;y \right),\\
& u_{2}(x)=e^{-y/2}y^{-\sigma}{}_{1}F_{1}\left( -\sigma+\frac{1}{2}-d;1-2\sigma;y \right) ,
\label{eq:MO2}
\end{aligned}
\end{equation}
where
\begin{equation}
y=2de^{-\gamma x}, \quad d^{2}=\frac{\Gamma_{0}}{\gamma^{2}}, \quad \sigma^{2}=\frac{\Gamma_{0}-\epsilon}{\gamma^{2}}, \quad W_{0}=2\sqrt{\Gamma_{0}-\epsilon}.
\label{eq:MO3}
\end{equation}
The physical energy eigenvalues are given by
\begin{equation}
E_{n}=\gamma\left[ (2n+1)\sqrt{\Gamma_{0}}-\gamma(n+1/2)^{2} \right], \quad n=0,1,\ldots N ,
\end{equation}
where $N$ is given by the floor function $N=\lfloor \frac{\sqrt{\Gamma_{0}}}{\gamma}-\frac{1}{2} \rfloor$. The related eigenfunctions can be recovered from (\ref{eq:MO2}) after substituting $E_n$ for $\epsilon$ and the appropriate boundary conditions. In Fig.~\ref{Fig3} we show the potential (\ref{eq:MO1}) and two of its supersymmetric partners for $\gamma=1$ and $\Gamma_{0}=4$. In such case, the initial potential admits two bound states with energy eigenvalues $E_{0}^{(0)}=7/4$ and $E_{1}^{(0)}=15/4$.  Notice that, besides the above energies, potentials $V_1(x)$ include the eigenvalue $E_0^{(1)}=\epsilon=1$ in their spectra. Moreover, they satisfy the condition of zero total area (\ref{zero}).

%%%%%%%%%%%%%
%%%%%%%%%%%%%

In summary, the method introduced in \cite{Ros15} and developed in \cite{Jai17,Ros18,Gar19} provides complex-valued potentials with all-real spectra that includes the parity-time-symmetric case as a particular result. The keystone of the approach relies on the solutions to the Ermakov equation (\ref{eq:FM6}) and the nonlinearity of the imaginary part of the superpotential (\ref{eq:LD1}). The latter permits to introduce the constraint (\ref{eq:FM8-1}) as an universal rule to choice the complex parameters that are required in the superposition (\ref{u}) to get properly defined complex potentials in supersymmetric quantum mechanics.

% ------------------------------------------------------------------------
\subsection*{Acknowledgment}
This research was funded by Consejo Nacional de Ciencia y Tecnolog\'ia (Mexico), grant number A1-S-24569, and by Instituto Polit\'ecnico Nacional (Mexico), project SIP20195981. K. Zelaya acknowledges the support from the Mathematical Physics Laboratory, Centre de Recherches Math\'ematiques, through a postdoctoral fellowship.

% ------------------------------------------------------------------------

% ------------------------------------------------------------------------
\end{document}